\documentclass[onecollarge,fleqn,runningheads]{svjour2}
\smartqed  
\usepackage{graphicx}
\journalname{Theoretical and Computational Fluid Dynamics}
\begin{document}
 
\title{A Comparative Study of Velocity Statistics of Hydrodynamic and Magnetohydrodynamic Turbulence}
%
%
\titlerunning{Velocity Statistics of HD and MHD Turbulence}


\author{Nicholas Hall \and G. Kowal \and A. Lazarian    \and
   Jungyeon Cho}

\authorrunning{Hall et al.} 

\institute{ Nicholas Hall, G. Kowal \& A. Lazarian \at 
   Astronomy Department, University of Wisconsin - Madison, Madison, 
   WI 53706
   \and
   Jungyeon Cho \at
   Department of Astronomy \& Space Science, Chungnam National      
   University, Daejeon, Korea \\
   jcho@cnu.ac.kr; (Phone) +82-42-821-5465; (Fax) +82-42-821-8891
}
\maketitle


\begin{abstract}
Turbulence in an incompressible fluid with and without a magnetic field as well as moderately compressible MHD turbulence are compared.  The results of three numerical simulation models in three dimensions of resolution up to $512^{3}$ are used for this purpose.  The compatibility of the spectra of all three models with the Kolmogorov spectrum is confirmed.  For the magnetohydrodynamic (MHD) models the probability distribution functions of the velocity components perpendicular to the external magnetic field are like the incompressible hydrodynamic (HD) model while those parallel to the field have a smaller range of velocities.  The probability distribution functions of the transverse velocity increments for the MHD models decline slower than the incompressible HD model.  The similarity of incompressible HD and both incompressible and compressible MHD turbulence persists over high order longitudinal structure function scaling exponents measured in the global reference frame as well as for motions perpendicular to the local mean field.  In these two frames the longitudinal scaling exponents of both MHD models seem to follow theoretical incompressible HD dissipation structure predictions while the transverse scaling exponents of the incompressible MHD model seem to follow the predictions for incompressible MHD.  In the local magnetic system the motions parallel to the local mean field for both MHD models are different from incompressible HD motions.  The fields of the MHD simulations are decomposed into Alfvenic, fast, and slow modes.  In the global reference frame and for motions perpendicular to the local mean field the Alfvenic mode is mostly responsible for the fact that the longitudinal components of both MHD models follow the incompressible HD model and the incompressible HD theoretical dissipation structure predictions.  In the global reference frame for the incompressible MHD model both the Alfvenic and slow modes seem to contribute significantly to the fact that the transverse components follow the incompressible MHD theoretical predictions.  For motions perpendicular to the magnetic field for the incompressible MHD model it appears that the slow mode contributes most to the fact that the transverse component seems to follow the incompressible MHD predictions.  



\end{abstract}

\section{Introduction}

Given that turbulent motions (velocities) are inherently stochastic, velocity field statistics are an essential tool for extracting its physical properties.  In 1941 Kolmogorov \cite{kol41} published his theory (hereafter K41) on the local velocity field structure of turbulence for incompressible hydrodynamic (hereafter IHD) fluids with high Reynolds numbers, in which he hypothesized the existence of homogeneity, isotropy, and an inertial range (hereafter K41).  The inertial range in the kinetic energy spectrum is a range in wave number, $k\sim1/l$ where $l$ is the separation length between two points often regarded as the eddie size, which lies between the typical large scales and the dissipative scales and is therefore independent of external input such as turbulence drive or output such as dissipation \cite{bis03}.  In this range the spectrum is also independent of viscosity and exhibits a power-law dependence of the form E(k)=K$\bar{\epsilon}^{-2/3}$$k^{-5/3}$, where K is the universal Kolmogorov constant and $\bar{\epsilon}$ is the average rate of energy dissipation per unit mass.   

It is known that the inclusion of an external magnetic field in the turbulence of an incompressible fluid brings about many substantial complications and can drastically modify the turbulent motions.  The inclusion of compressibility also causes significant changes in the properties of turbulence.  Gotoh, Fukayama, \& Nakano \cite{got02} thoroughly and successfully investigated the velocity field statistics of high-resolution direct numerical simulations for IHD turbulence.  The purpose of this paper is to address the question, are the velocity field statistics completely different from IHD when a magnetic field and compressibility are included?  We extend the analysis of Gotoh et al. \cite{got02} to incompressible magnetohydrodynamic (hereafter IMHD) and compressible magnetohydrodynamic (hereafter CMHD) turbulence in an attempt to answer this question.  This is considered mostly through the use of high order scaling exponents of the velocity structure functions.  

Emphasis is also placed on the differences between the scaling exponents of longitudinal and transverse structure functions.  For high order scaling exponents the findings of Gotoh et al. \cite{got02} are confirmed for structure functions measured in the global reference frame for IHD turbulence.  Scaling exponents are then explored for IMHD and CMHD turbulence in the global frame and the local frame, defined by the local mean magnetic field.  The scaling exponents are then decomposed into Alfvenic, fast, and slow modes and the contributions of individual modes are explored.  The paper is organized as follows.  Sec. 2 presents the numerical aspects of the three numerical simulation models considered (IHD, IMHD, and CMHD).  Sec. 3 contains analysis of the kinetic energy spectra of the three models.  In Sec. 4 probability distribution functions are considered.  Structure functions are explored in Sec. 5.  Sec. 6 presents scaling exponents of the structure functions.  In Sec. 7 the scaling exponents are decomposed into the three MHD modes.  Finally, Sec. 8 includes a discussion and summary.

\section{Simulation Specifics}

The IHD data used in this study came from simulations with a grid size of $256^3$.  For both MHD models, a high resolution grid size of $512^3$ was used and the external magnetic field is in the positive x-direction.

\subsection{\label{sec:level2}Incompressible HD and MHD}
The IMHD code described below is used for the IHD model with $B$ set to zero.  We have calculated the time evolution of incompressible magnetic turbulence subject to a random driving force per unit mass.  We have adopted a pseudospectral code to solve the incompressible MHD equations in a periodic box of size $2\pi$:
\begin{equation}
\frac{\partial {\bf v} }{\partial t} = (\nabla \times {\bf v}) \times {\bf v}
      -(\nabla \times {\bf B})
        \times {\bf B} + \nu \nabla^{2} {\bf v} + {\bf f} + \nabla P' ,
        \label{veq}
\end{equation}
\begin{equation}
\frac{\partial {\bf B}}{\partial t}=
     \nabla \times ({\bf v} \times{\bf B}) + \eta \nabla^{2} {\bf B} ,
     \label{beq}
\end{equation}
\begin{equation}
      \nabla \cdot {\bf v} =\nabla \cdot {\bf B}= 0,
\end{equation}
where $\bf{f}$ is a random driving force, $P'\equiv P/\rho + {\bf v}\cdot {\bf v}/2$, ${\bf v}$ is the velocity,
and ${\bf B}$ is magnetic field divided by $(4\pi \rho)^{1/2}$.  In this representation, ${\bf v}$ can be viewed as the velocity measured in units of the r.m.s. velocity, v, of the system and ${\bf B}$ as the Alfv\'en speed in the same units.  The time $t$ is in units of the large eddy turnover time ($\sim L/v$) and the length in units of $L$, the inverse wavenumber of the fundamental box mode.  In this system of units, the viscosity $\nu$ and magnetic diffusivity $\eta$ are the inverse of the kinetic and magnetic Reynolds numbers respectively.  The magnetic field consists of the uniform background field and a
fluctuating field: ${\bf B}= {\bf B}_0 + {\bf b}$.  We use 21 forcing components with $2\leq k \leq \sqrt{12}$, where wavenumber $k$ is in units of $L^{-1}$.  Each forcing component has correlation time of one.
The peak of energy injection occurs at $k\approx 2.5 $.  The amplitudes of the forcing components are tuned to ensure $v \approx 1$.  We use exactly the same forcing terms
for all simulations.  The Alfv\'en velocity of the uniform background field, $B_0$, is set to 1.  We consider only cases where viscosity is equal to magnetic diffusivity:
\begin{equation}
  \nu = \eta.
\end{equation}
In pseudo spectral methods, the temporal evolution of equations (\ref{veq}) and (\ref{beq}) are followed in Fourier space.  To obtain the Fourier components of nonlinear terms, we first calculate them in real space, and transform back into Fourier space.  The average kinetic helicity in these simulations is not zero. However, previous tests have shown that our results are insensitive to the value of the 
kinetic helicity.  In incompressible fluid, $P'$ is not an independent variable.  We use an appropriate projection operator to calculate $\nabla P'$ term in {}Fourier space and also to enforce divergence-free condition ($\nabla \cdot {\bf v} =\nabla \cdot {\bf B}= 0$).  We use up to $256^3$ collocation points.  We use an integration factor technique for kinetic and magnetic dissipation terms and a leap-frog method for nonlinear terms.  We eliminate the $2\Delta t$ oscillation of the leap-frog method by using
an appropriate average.  At $t=0$, the magnetic field has only its uniform component and the velocity field is restricted to the range $2\leq k \leq 4$ in wavevector space.

Hyperviscosity and hyperdiffusivity are used for the dissipation terms.  The power of hyperviscosity is set to 6, so that the dissipation term in the above equation is replaced with
\begin{equation}
 -\nu_6 (\nabla^2)^6 {\bf v},
\end{equation}
where $\nu_6$ is determined from the condition $\nu_h (N/2)^{2h} \Delta t \approx 0.5$ (see \cite{bor96}).  Here $\Delta t$ is the time step and $N$ is the number of grid points in each direction.  The same expression is used for the magnetic dissipation term.  Cho \& Lazarian \cite{cl02} can be referenced for more details on the above discussion.  

\subsection{\label{sec:level2}Compressible MHD}

We use a third-order accurate hybrid essentially non-oscillatory (ENO) scheme (see \cite{cho02}) to solve the ideal isothermal MHD equations in a periodic box:
\begin{eqnarray}
{\partial \rho    }/{\partial t} + \nabla \cdot (\rho {\bf v}) =0,  \\
{\partial {\bf v} }/{\partial t} + {\bf v}\cdot \nabla {\bf v} 
   +  \rho^{-1}  \nabla(a^2\rho)
   - (\nabla \times {\bf B})\times {\bf B}/4\pi \rho ={\bf f},  \\
{\partial {\bf B}}/{\partial t} -
     \nabla \times ({\bf v} \times{\bf B}) =0, 
\end{eqnarray}
with $ \nabla \cdot {\bf B}= 0$ and an isothermal equation of state.  Here $\bf{f}$ is a random large-scale driving force, $\rho$ is density, ${\bf v}$ is the velocity, and ${\bf B}$ is magnetic field.  The rms velocity $\delta V$ is maintained to be approximately unity (in fact $\delta V \sim 0.7$), so that ${\bf v}$ can be viewed as the velocity measured in units of the r.m.s. velocity of the system and ${\bf B}/\sqrt{4 \pi \rho}$ as the Alfv\'{e}n velocity in the same units.  The time $t$ is in units of the large eddy turnover time ($\sim L/\delta V$) and the length in units of $L$, the scale of the energy injection.  The magnetic field consists of the uniform background field and a fluctuating field: ${\bf B}= {\bf B}_0 + {\bf b}$.  For our calculations we assume that $B_0/\sqrt{4 \pi \rho} \sim \delta B/\sqrt{4 \pi \rho} \sim \delta V$.  The sound speed is one therefore the resulting Mach number and plasma $\beta$ are of order unity. 
As is the case with other finite-difference numerical schemes, 
our compressible MHD code inherits numerical viscosity (and
diffusion), which results in energy dissipation on small scales.  
 See \cite{cl03} for details on the above discussion.

\section{\label{sec:level1}Kinetic Energy Spectrum}

\begin{figure*}
\includegraphics{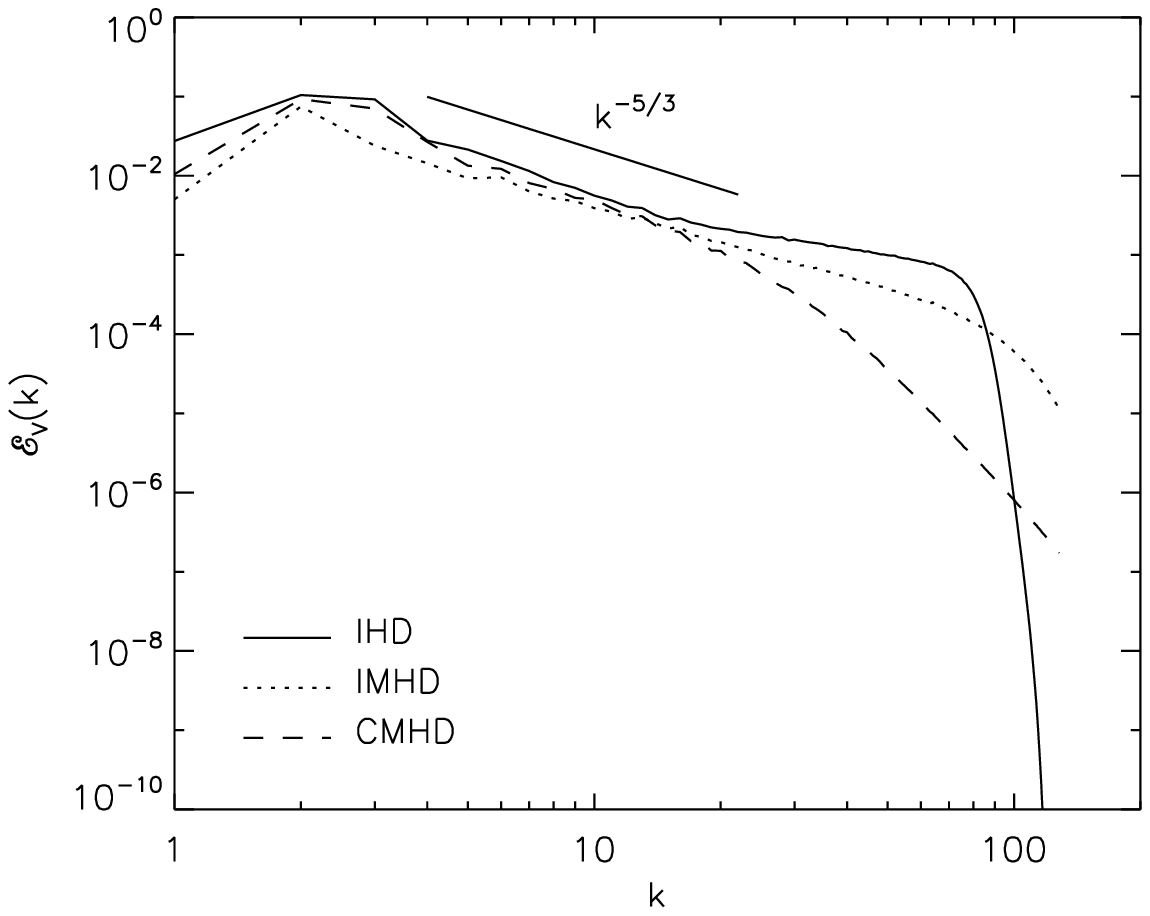}
\caption{Kinetic energy spectra for the three models as well as Kolmogorov's power-law dependence in the inertial range for comparison.\label{spec}}
\end{figure*}

The three-dimensional kinetic energy spectrum for each model, taken 
after turbulence has reached a statistically stationary state, is shown in Figure \ref{spec}.  The inertial range of the spectrum is of most interest for this study.  The shape of the curves at wave numbers less than those of the inertial range depends on the energy at the injection scale and how the turbulent motions are excited.  The shape of the curves at wave numbers greater than those of the inertial range is affected significantly by viscosity, the bottleneck effect \cite{fal94}, and the resolution of the model.  Despite the fact that the Kolmogorov phenomenology, E(k)$\propto$$k^{-5/3}$, is derived for IHD, in the inertial range all three curves demonstrate a very similar power law dependence, suggesting that the energy transfer process is consistent over this range of wave numbers.  This may be surprising considering the many changes resulting from the inclusion of a magnetic field, such as an Alfvenic mode propagating parallel to the magnetic field and elongation of the structures in the velocity field.  The inclusion of compressibility introduces significant changes as well, such as shocks.  Goldreich \& Sridhar \cite{gol95} provide an explanation of this behavior.  See \cite{cl05} also for a review.  Deviations from the $5/3$ law have become an issue of debate (see \cite{bol06,ber06}, and references therein).  We do not address this issue here.

\section{\label{sec:level1}Probability Distribution Functions}

\begin{figure*}
\includegraphics{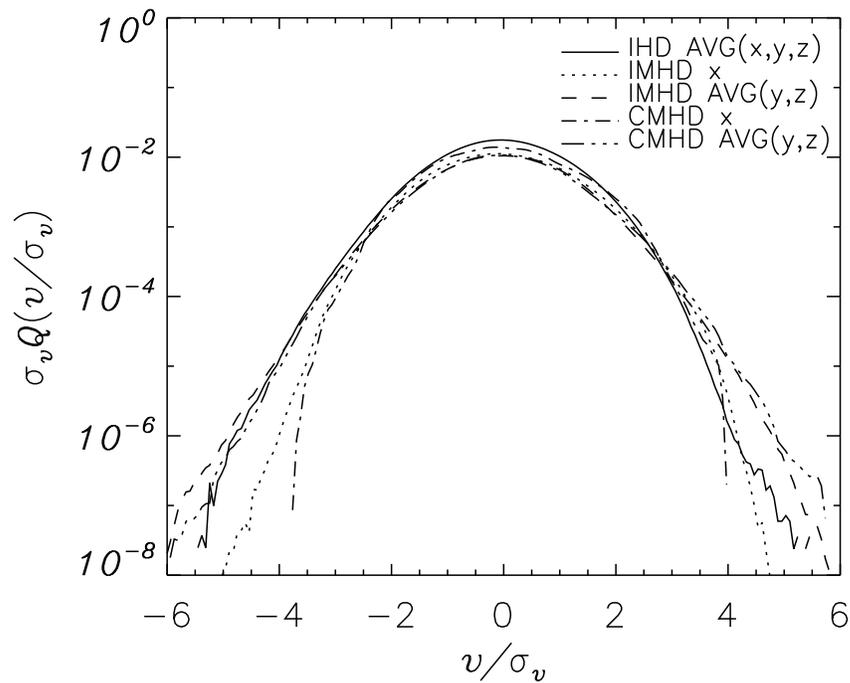}
\caption{PDF's of the velocity field parallel (x) and perpendicular (y \& z) to $B_{ext}$ for the three models.\label{pdf}}
\end{figure*}

The probability distribution function (hereafter PDF) is used to obtain information about one-point velocity statistics (see \cite{elmS03} for
detailed discussions on astrophysical applications).  The single-point PDF's of different components of the velocity field for each model shown in Figure \ref{pdf} are histograms, normalized to the grid size, of the values of the velocity at each point.  For the IHD model the x, y, and z components were averaged.  For the two MHD models, the histograms were separated into components parallel (x direction) and perpendicular (y \& z directions) to the external magnetic field, ($B_{ext}$); therefore, the PDF's for the x components are presented separately and the y and z components are averaged.

The PDF for the IHD model is somewhat asymmetric.  Any asymmetry or skewness in the PDF's should disappear when averaged over multiple time steps \cite{got02}.  The PDF's of the MHD simulations parallel to $B_{ext}$ are similar to each other, fairly symmetric, wider toward the top than a gaussian, and at large amplitudes decay faster than a gaussian.  Their slopes are the steepest of the PDF's presented and given that the area under the curve must be the same, their ranges of velocities are smaller.  For IMHD turbulence the divergence of the velocity vector is zero meaning that there are no sources, sinks, or shocks and that only eddies can exist.  This is not the case for CMHD simulations in which shocks are produced at high velocities which dissipate the energy perhaps contributing to the PDF of the parallel CMHD model having the least high velocity points.

The PDF's of the IMHD and CMHD models perpendicular to $B_{ext}$ approximately follow each other and are even more similar to each other than the parallel components.  They are very symmetric and gaussian-like, have the most shallow slopes of all, and therefore have the most high velocity points.  

When an external magnetic field is introduced into the turbulent motions, an Alfvenic mode is produced with perturbations perpendicular to the direction of the field.  This additional contribution to the velocity is responsible for the higher velocities found in the perpendicular PDF's.  The additional kinetic energy given to the perpendicular direction explains the separation of these two components seen in the top of Figure \ref{pdf}.  This appears to be equally valid for both positive and negative values of the velocity.

\begin{figure*}
\includegraphics{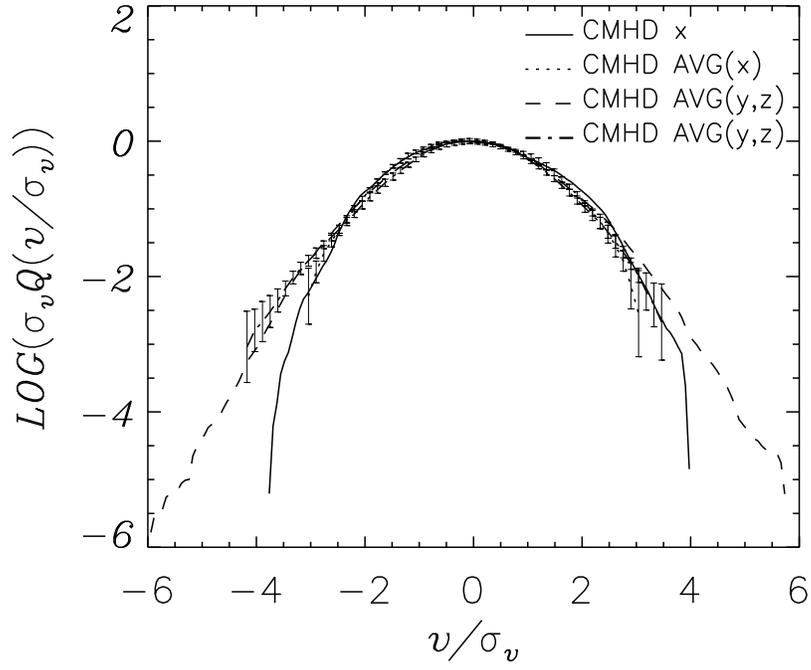}
\caption{PDF's, normalized to match their maxima, of the velocity field parallel and perpendicular to $B_{ext}$ for the one time step of the CMHD simulation used throughout the paper and the average over 10 time steps of a 256 resolution simulation with error bars showing expected random fluctuations.
\label{pdf_error}}
\end{figure*}

An average over 10 time steps of the velocity PDF's for a CMHD simulation with 256 resolution was performed.  The 10 parallel time steps were averaged as well as the 20 perpendicular time steps, 10 for y and 10 for z.  These PDF's are plotted in Figure \ref{pdf_error} along with the CMHD PDF's mentioned above.  The PDF's are normalized to match their maxima in order to better compare their tails.  The parallel and perpendicular PDF's for the single time step lie sufficiently close to the expected random errors obtained from the PDF's averaged over multiple time steps to justify using a single time step and averaging the perpendicular components.

\begin{figure*}
\includegraphics{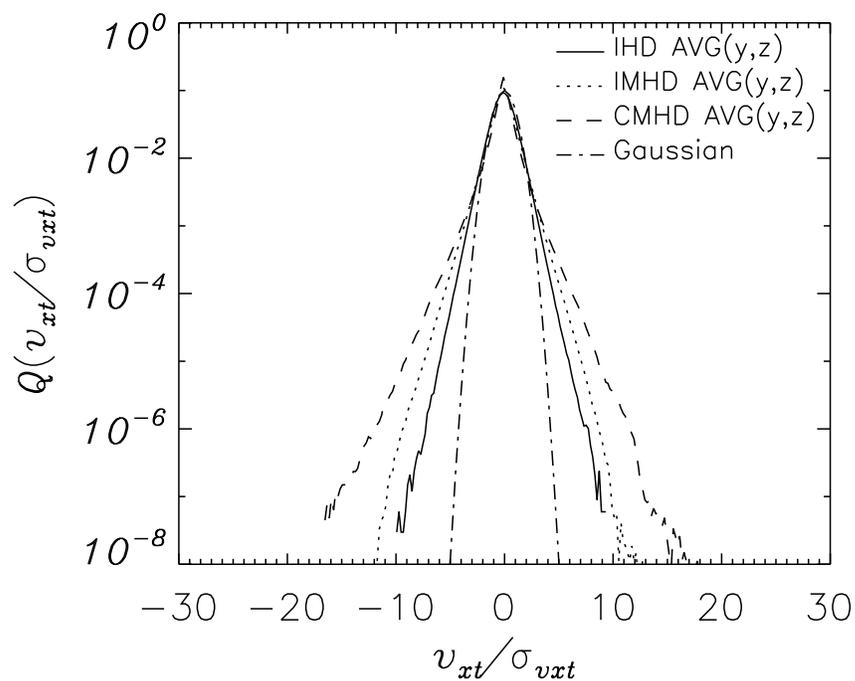}
\caption{PDF's of the transverse increments of the x component of the velocity field taken along y and z averaged together for the three models. 
 Note that the mean magnetic field is parallel
to the x-axis in the MHD cases.
\label{pdf_derx}}
\end{figure*}

\begin{figure*}
\includegraphics{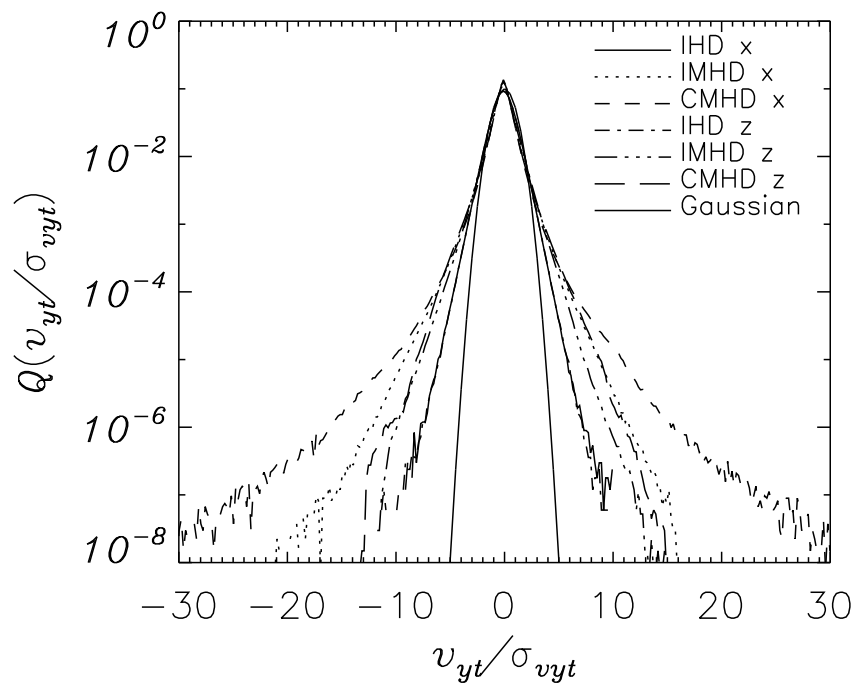}
\caption{PDF's of the transverse increments of the y component of the velocity field taken along x and z presented separately for the three models.
\label{pdf_dery}}
\end{figure*}

Figures \ref{pdf_derx} and \ref{pdf_dery} show PDF's of the transverse increments of the velocity field which are histograms, normalized to the grid size, of the values of the difference between the velocity field at each point and the field at an adjacent point along the direction specified.  Positive values suggests that $v$ is increasing in the positive direction while negative values suggest that it is decreasing.  In Figure \ref{pdf_derx} the increments of the x-component of the velocity field taken along the y and z directions, parallel to $B_{ext}$ for the MHD models, are averaged.  Figure \ref{pdf_dery} shows the increments of the y-component along the x and z directions.  The results are not averaged given that one is parallel to $B_{ext}$ while the other is perpendicular.  

In both Figures, all three models demonstrate near perfect symmetry.  This is true despite the orientation with respect to $B_{ext}$.  In Figure \ref{pdf_dery} as expected for the IHD model, x and z are nearly identical because there is no magnetic field to bring about differences.  On the other hand for the MHD models, x extends to higher values of the velocity increment than z.

Because of the fact that the difference in range between the x components for the IMHD and CMHD models is much greater than the difference in range between the z components for the IMHD and CMHD models, it is possible that most of the increase in range in going from IMHD models to CMHD models by introducing compressibility is coming from the components parallel to $B_{ext}$.  This is supported by the fact that for the PDF's in Figure \ref{pdf} the perpendicular components follow each other closely while the parallel components are quite different from each other.

\section{Structure Functions}

Structure functions are used to obtain information about two-point velocity statistics.  Structure functions describe the distribution of velocity structures of size $l$ and are defined as
\begin{equation}
 S^{(p)}(l)=\langle | \delta v (\vec{l}) | ^p \rangle,
 \label{eqn:str_func}
\end{equation}
where $\delta v$ is a velocity increment and $\langle ... \rangle$ represents ensemble averaging.  Structure functions are defined with the absolute value of the velocity increment to ensure their positivity for all orders.  Structure functions are separated into a longitudinal component naturally defined by the velocity increment 
\begin{equation}
 \delta v_L(l) = [\vec{v}(\vec{x} + \vec{l}) - \vec{v}
(\vec{x})]\cdot\hat{l},
 \label{eqn:vel_incr}
\end{equation}
and transverse components $\delta v_T (l)$ chosen randomly in the plane perpendicular to the longitudinal component.  When higher order structure functions are used the PDF tails contribute more, where as you can see in Figure \ref{pdf_error} the errors are much larger.  

For each separation length $l$ from 1 to 128,  $\delta v$ is calculated at 100,000 points randomly distributed throughout the entire periodic velocity field.  In the global reference frame the direction of $\hat{l}$, or the direction of the longitudinal structure function, is also random.  The structure functions for the IHD model are calculated in the global reference frame.  For the MHD models $\hat{l}$ can be chosen randomly (global frame), parallel to the local mean magnetic field, or perpendicular to the local mean field.  

\begin{figure*}
\includegraphics{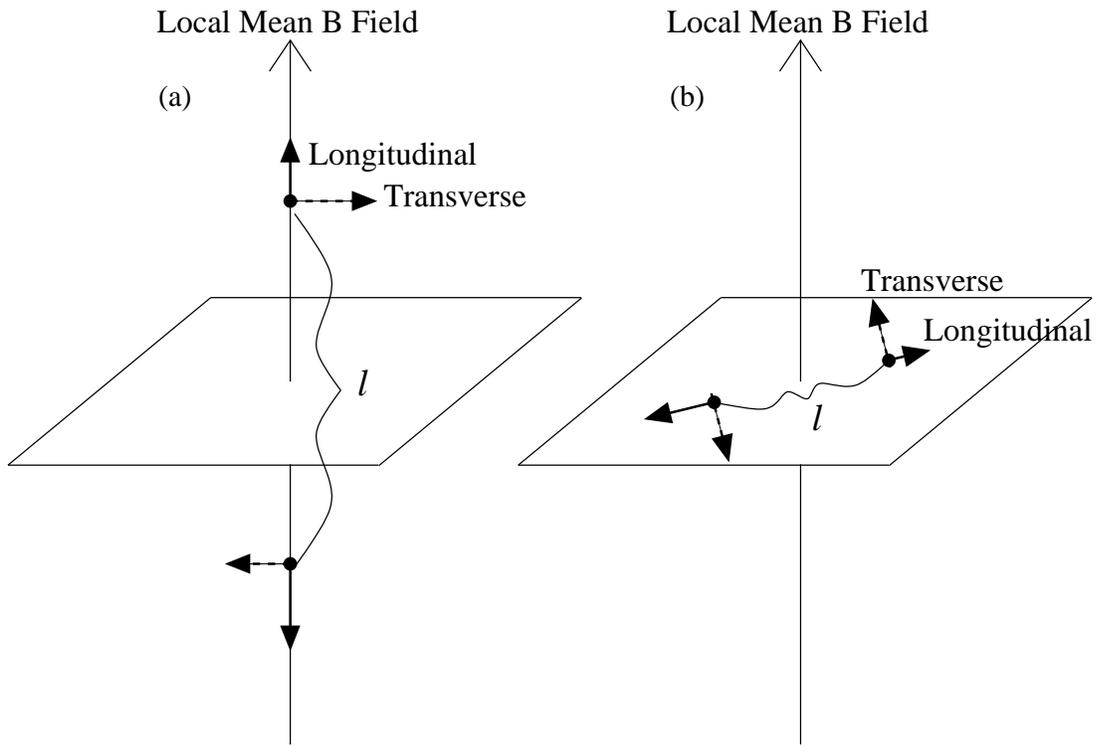}
\caption{For (a) points are chosen along the local mean field, hereafter referred to as local parallel.  For (b) points are chosen in the plane perpendicular to the local mean field so that both the longitudinal and the transverse structure functions lie in this plane, hereafter referred to as local perpendicular.}
\label{local_explan}
\end{figure*}

It is well known that, 
for MHD turbulence in a strongly magnetized medium
energy cascade occurs mainly in the direction
perpendicular to the mean field in the Fourier space, which makes
eddies elongated \cite{Mon81,She83,Mon95,Mat98} along 
the global mean magnetic field.
However, according to \cite{cho00a} and \cite{cho02}, 
for MHD turbulence in a strongly magnetized medium eddies are not aligned along the direction of the total mean field but rather are aligned along the local mean field.  This being the case, the local mean field is the physically relevant background for the eddie dynamics.  The local frame is separated into two types which are represented in Figure \ref{local_explan}.  In (a) the points from which the structure functions are calculated are chosen along the local mean field, this configuration will be referred to as local parallel.  In (b) the points are chosen in the plane perpendicular to the local mean field, this configuration will be referred to as local perpendicular.
 
\begin{figure*}
\includegraphics{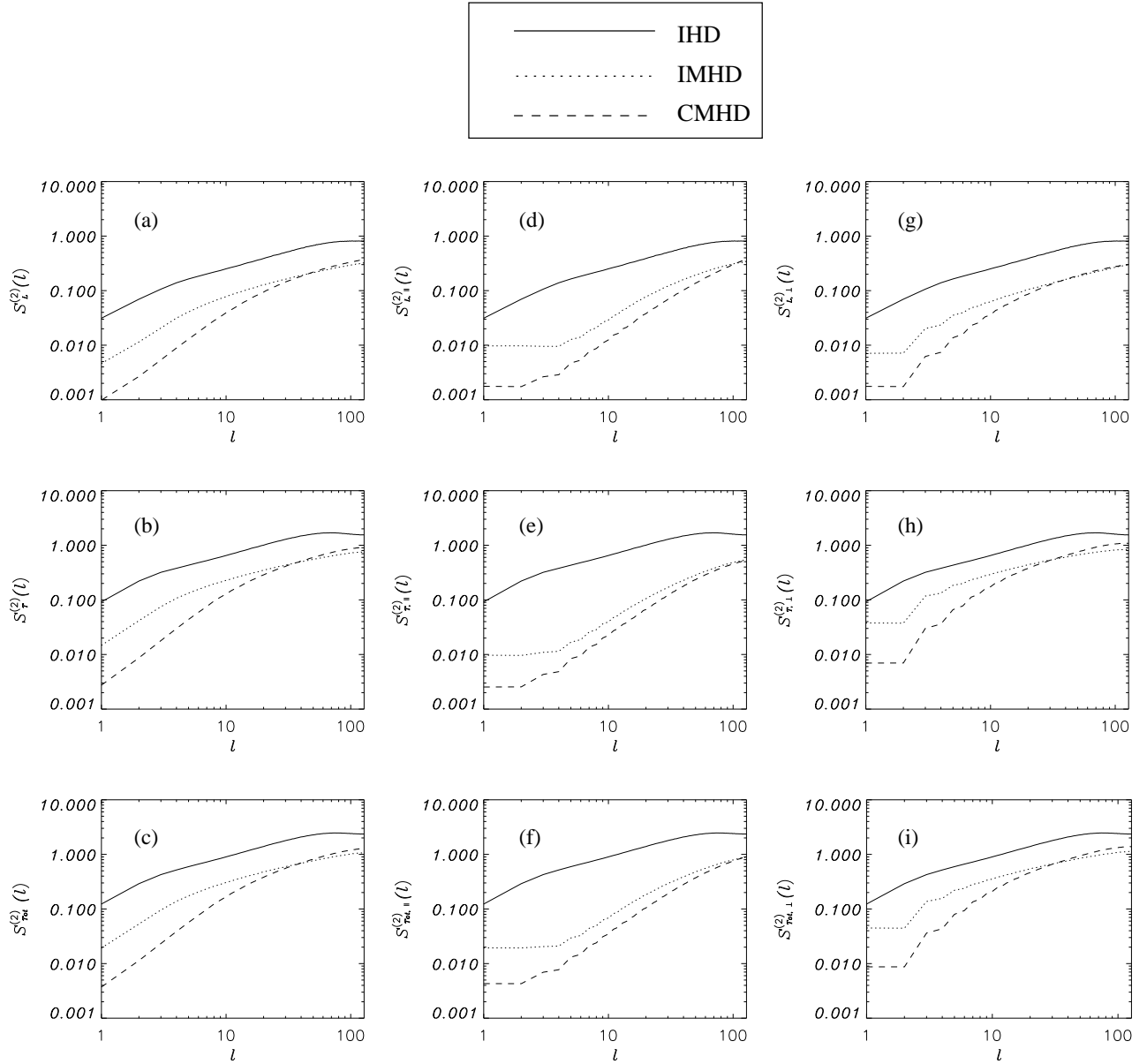}
\caption{Second order structure functions of the velocity field for the three models.  The top row are longitudinal, the middle row are transverse, and the bottom row are total structure functions.  The first column is in the global reference frame, the middle row is the local parallel frame, and the bottom row is the local perpendicular frame.  In the local frames the HD structure functions from the global frame are shown for comparison.
\label{sfs}}
\end{figure*}

Figure \ref{sfs} shows the second order longitudinal, transverse, and total structure functions of the velocity field.  The column on the left is the global frame, the middle column is the local parallel frame, and the column on the right is the local perpendicular frame.  In the local frames the structure functions of the IHD model calculated in the global frame are shown for comparison.

The inertial range of the turbulent motions manifests itself in the structure functions as the range of $l$ over which the slope is constant in a log-log plot.  The first maximum of a structure function is related to the size of the largest structures in the corresponding field, the radius of the largest eddies for example.  It will be found at the separation length at which the largest $\delta v$ is observed.

We confirm the result of \cite{got02} that in the global frame for IHD turbulence the values of the transverse structure functions are greater than those of the longitudinal.  In extending our analysis to IMHD and CMHD we find that the structure functions of both MHD models in the global and local perpendicular reference frames exhibit this behavior as well.  For each MHD model in the local parallel reference frame the values of the longitudinal and transverse structure functions are very similar.  In the global and local parallel frames the IMHD structure functions are greater than those of the CMHD model at lower values of the separation length where they appear more like the IHD structure functions in the global frame.  In summary, the MHD models in the global frame and in the plane perpendicular to the local mean field are significantly more similar to the IHD structure functions than the structure functions of the MHD models in the local parallel frame.  Also, the shapes of the structure functions in the global and local parallel frames for the IMHD model are more like that of the IHD model than the CMHD model is.

\section{Scaling Exponents}

Self-similarity or scale-invariance implies that fluid turbulence at larger scales can be reproduced by the magnification of turbulent motions at smaller scales.  At the dissipation scales self similarity fails as turbulence forms non-Gaussian dissipation structures \cite{bis03}.  Self-similarity is even an approximation in the inertial range where turbulent motions become increasingly sparse in space and time at smaller scales, a property known as intermittency.  In 1962 Kolmogorov published a modification of his 1941 theory including the effects of intermittency or scale-variance \cite{kol62}.  One way to investigate intermittency is by studying the scaling exponents of velocity fluctuations in the inertial range, given that for a scale-invariant flow the scaling exponents are linear functions of the order of the structure functions.  

\begin{figure*}
\includegraphics{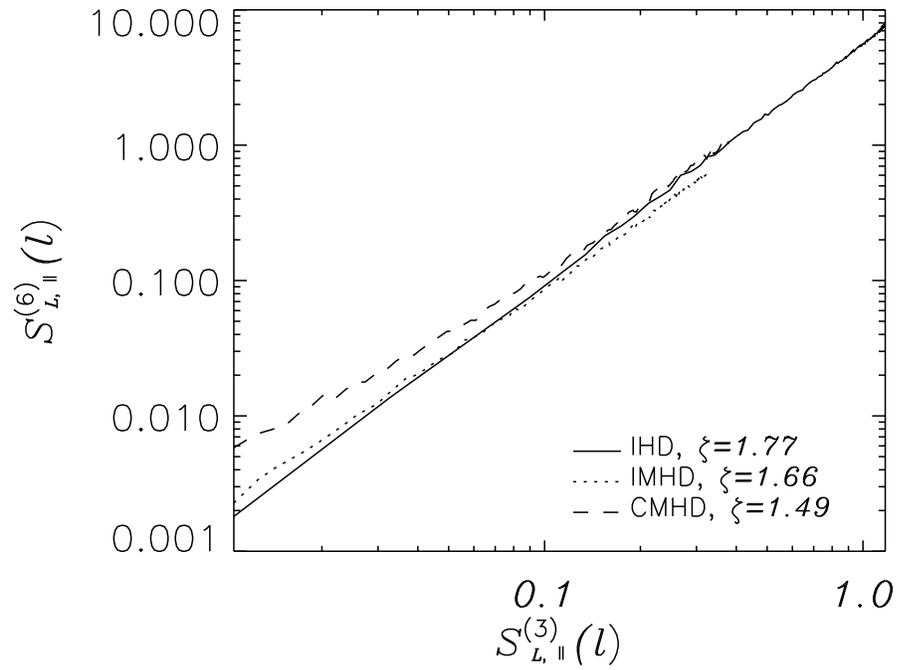}
\caption{The sixth order longitudinal structure function as a function of the third order longitudinal structure function both calculated in the local parallel frame for the three models, along with their values of scaling exponents obtained from a minimum chi-square linear fit.  Compare to Figure \ref{ses_loc_para} (a).
\label{selfsim}}
\end{figure*}

Despite the fact that the origin is not yet fully understood, the scaling behavior of structure functions is very useful and interesting (see \cite{elmS03} for a comprehensive discussion on astrophysical
situations;
see also \cite{laz06}).  Within the inertial range, a structure function of arbitrary order scales with the separation length $l$ as $S^{(p)}(l)\propto{l^{\zeta(p)}}$. From this scaling relation and a few transformations self-similarity between two structure functions of arbitrary orders, $p$ and $p'$, can be obtained: 
\begin{equation}
 S^{(p)}(l) \propto{[S^{(p')}(l)]^{\zeta(p)/\zeta(p')}},
 \label{eqn:sf_relation}
\end{equation}
demonstrating how the structure functions $S^{(p)}(l)$ and $S^{(p')}(l)$ are related (see \cite{bis03}).  It also brings to light the possibility of using a structure function of an arbitrary order $p'$ as a normalization.  In Figure \ref{selfsim} the sixth order longitudinal structure function is plotted as a function of the third order longitudinal structure function where $\hat{l}$ is parallel to the local mean magnetic field, demonstrating how self similarity can be used to obtain scaling exponents.  A minimum chi-square linear fit was used to determine $\zeta(6)$.  The scaling relation $S^{(p)}(l) \propto{l^{\zeta(p)}}$ in the inertial range has been derived analytically in Kolmogorov phenomenology yielding $\zeta(p)=p/3$.  For this reason the third moment is used as the normalization.  If the effects of dissipation on the structure functions are independent of the order one may conclude that not only the third order moment but any moment $S^{(p')}(l) \propto{l^{\zeta(p')}}$ can be used \cite{bis03}.   

She \& Leveque \cite{she94} proposed a scaling relation that contains three parameters (see also \cite{pol95,mul00,bol02}): $g$ related to the scaling $\delta v\sim l^{\frac{1}{g}}$, $x$ related to the energy cascade rate $t^{-1}\sim l^{-x}$, and $C$ the codimension of the dissipative structures:
\begin{equation}
\zeta(p) = \frac{p}{g}(1-x)+C[1-(1-\frac{x}{C})^{\frac{p}{g}}].
\end{equation}
For IHD turbulence using $g=3$, $x=\frac{2}{3}$, and $C=2$ She \& Leveque \cite{she94} obtained
\begin{equation}
\zeta^{SL1}(p)=\frac{p}{9}+2[1-(\frac{2}{3})^{\frac{p}{3}}],
\end{equation}
suggesting that dissipation occurs over one-dimensional structures, such as filaments or vortices (hereafter SL1).  For three-dimensional IMHD turbulence Muller \& Biskamp \cite{mul00} (see also \cite{mul03}) in extending the She \& Leveque \cite{she94} model to IMHD turbulence and using $g=3$, $x=\frac{2}{3}$, and $C=1$ obtained
\begin{equation}
\zeta^{SL2}(p)=\frac{p}{9}+1-(\frac{1}{3})^{\frac{p}{3}}
\end{equation}
assuming the dissipative structures to be two-dimensional, such as sheets (hereafter SL2).  Finally, for CMHD turbulence Padoan et al. \cite{pad04} showed that the velocity field for subsonic MHD turbulence follows SL1, while for highly supersonic MHD turbulence it follows SL2.

\begin{figure*}
\includegraphics{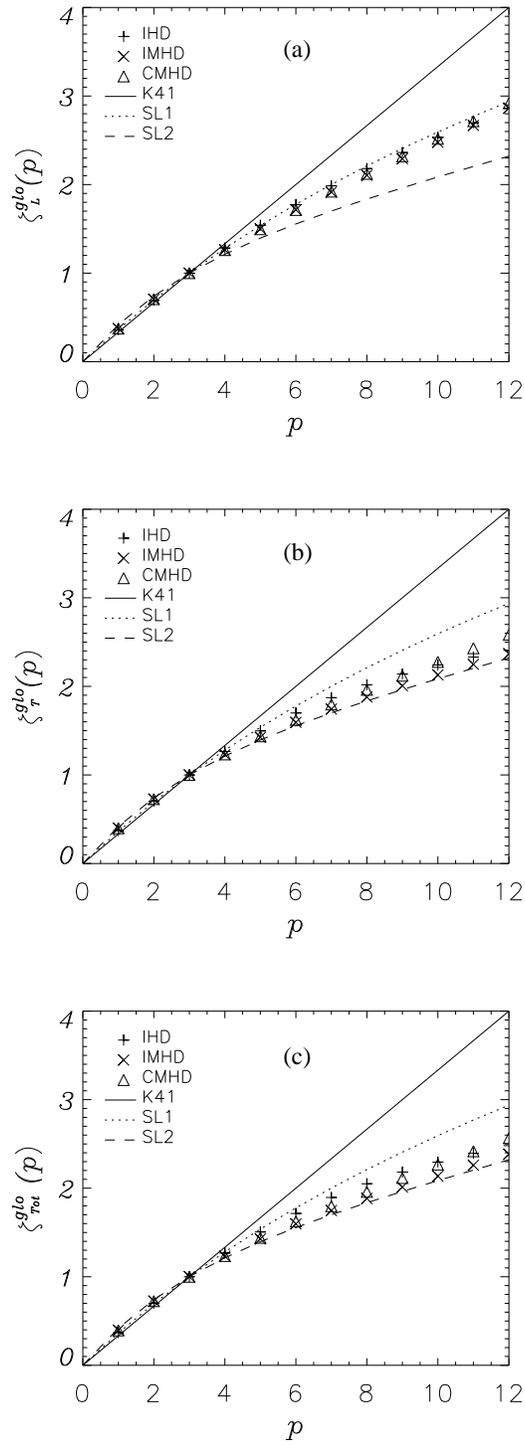}
\caption{Scaling exponents, normalized to the third order, as a function of the order of the structure functions in the global reference frame for the three models as well as K41, SL1, and SL2 ((a) - longitudinal, (b) - transverse, and (c) - total).  
\label{ses_global}}
\end{figure*}

\begin{figure*}
\includegraphics{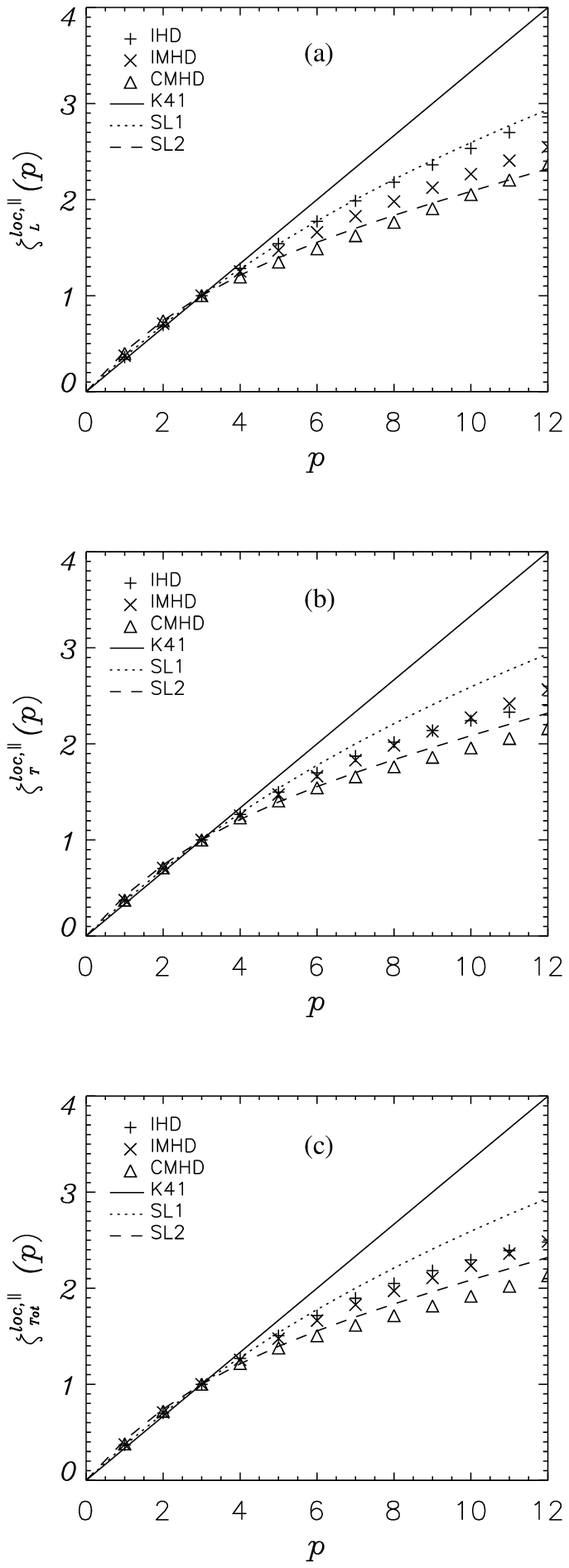}
\caption{Scaling exponents, normalized to the third order, as a function of the order of the structure functions in the local parallel reference frame for the three models as well as K41, SL1, and SL2 ((a) - longitudinal, (b) - transverse, and (c) - total).  For the IHD model the global scaling exponents are shown for comparison.
\label{ses_loc_para}}
\end{figure*}

\begin{figure*}
\includegraphics{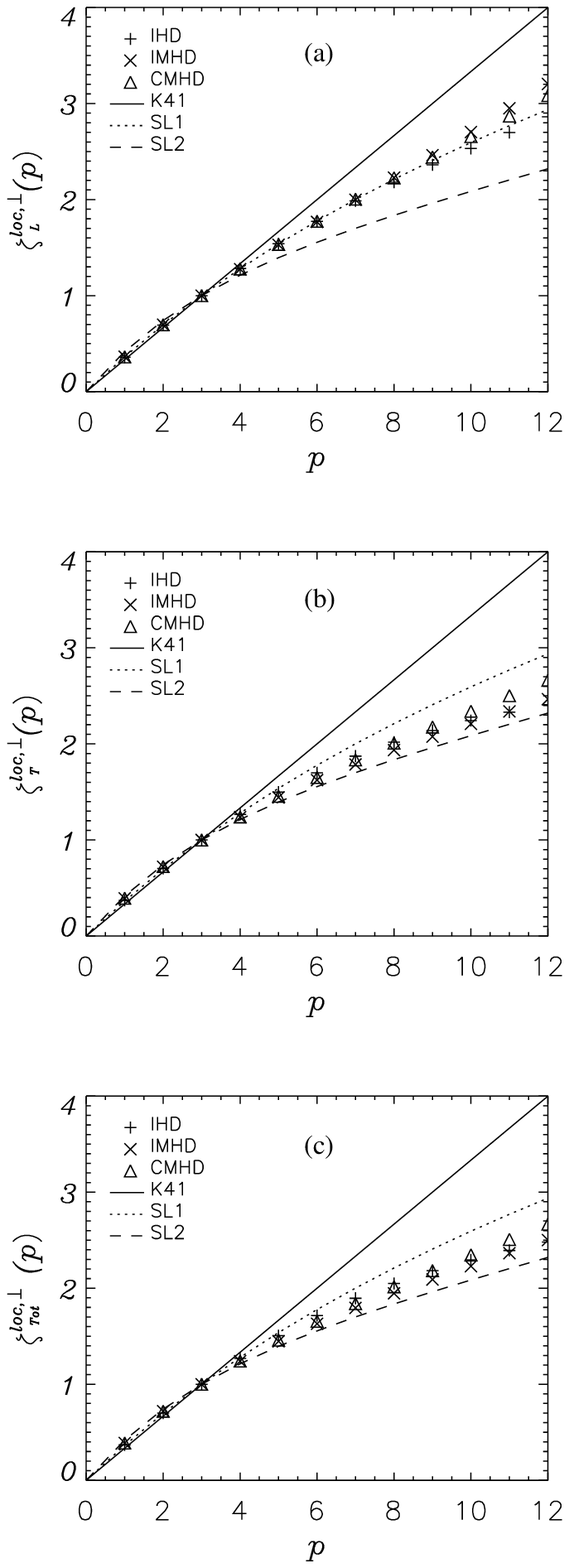}
\caption{Scaling exponents, normalized to the third order, as a function of the order of the structure functions in the local perpendicular reference frame for the three models as well as K41, SL1, and SL2 ((a) - longitudinal, (b) - transverse, and (c) - total).  For the IHD model the global scaling exponents are shown for comparison.
\label{ses_loc_perp}}
\end{figure*}

Figures \ref{ses_global} through \ref{ses_loc_perp} contain the scaling exponents as a function of the order of the longitudinal, transverse, and total structure functions normalized to the third order.  For the MHD models the Figures are organized as follows.  In Figure \ref{ses_global} the global reference frame is used, in Figure \ref{ses_loc_para} the local parallel reference frame is used, and in Figure \ref{ses_loc_perp} the local perpendicular reference frame is used.  For the IHD model the global scaling exponents are also presented in Figures \ref{ses_loc_para} and \ref{ses_loc_perp} for comparison, given that the local reference frame has no meaning without a magnetic field.  Despite the fact that the error in the scaling exponents increases with order from the increasing fluctuation in the structure functions, emphasis is placed on high order structure functions ($p>4$) given that at low orders the theoretical predictions considered here differentiate themselves very little.

Beginning with Figure \ref{ses_global} in the global reference frame the longitudinal scaling exponents of all three models follow SL1 quite well.  In the transverse case although the slope of the IHD model initially follows SL1 it begins to flatten at around $p=7$.  Its shape is in agreement with the results of \cite{got02}.  The IMHD model follows SL2 quite well while the CMHD model is between the two.  The IHD results mentioned here agree with those obtained by \cite{got02} in that for $p>4$ the longitudinal scaling exponents fall slightly below SL1 while the transverse exponents are significantly lower.  For all three models, despite the fact that the longitudinal scaling exponents follow SL1 much better than the transverse follow SL2, the scaling exponents calculated from the total structure functions are much more SL2-like than SL1-like for all three models.  Their behavior is very similar to the transverse case, suggesting that the transverse scaling exponents have a significantly larger amplitude than the longitudinal, which can be seen to be true in Figure \ref{sfs}.  The same holds for Figures \ref{ses_loc_para} and \ref{ses_loc_perp}.

In Figure \ref{ses_loc_para} in the local parallel reference frame for the longitudinal component, parallel to the local mean magnetic field in this case, the IMHD model is between SL1 and SL2 while the CMHD model mostly follows SL2.  For the transverse component, perpendicular to the local mean magnetic field in this case, the IMHD model is between SL1 and SL2 while the CMHD model follows SL2.  

In Figure \ref{ses_loc_perp} in the local perpendicular reference frame for the longitudinal component, perpendicular to the local mean magnetic field in this case, the best agreement to SL1 is found for all three models at all values of $p$.  For the transverse component, also perpendicular to the local mean magnetic field in this case, we do not find a similar agreement; the IMHD model is more SL2-like while the CMHD model remains in between.    

We confirm the results of \cite{got02} that for IHD turbulence the transverse scaling exponents, $\zeta_T(p)$ are less than the longitudinal $\zeta_L(p)$ for orders greater than four.  In extending the analysis to incompressible and compressible MHD turbulence we find that in the global reference frame and in the local perpendicular reference frame this behavior continues.  This is not necessarily the case in the local parallel reference frame.  This difference is probably related to the fact that in the local parallel frame the longitudinal component necessarily probes the direction parallel to the local mean field while the transverse component probes the plane perpendicular to the local mean field, thus incorporating a consistent difference between the two components that does not exist in the other two frames.

The most conspicuous SL1 cases above are both longitudinal, namely $\zeta^{glo}_{L}(p)$ and $\zeta^{loc, \bot}_{L}(p)$ in which all models follow SL1 suggesting physically that the dissipation structures are one dimensional hydrodynamic-like vortices.  The most conspicuous SL2 cases are both the transverse cases corresponding to the longitudinal SL1 cases above, namely  $\zeta^{glo}_{T}(p)$ and $\zeta^{loc, \bot}_{T}(p)$ in which the IMHD model mostly follows SL2 suggesting physically that the dissipation structures are two dimensional sheets dominated by MHD processes.  It is possible and perhaps more likely that the dissipation structures only appear one or two dimensional in slices of the three dimensional MHD turbulence \cite{cho03a}.  The cases in which the scaling exponents fall in between SL1 and SL2 may be some mixture of the two physical pictures.  Also, in the global reference frame and in the local perpendicular reference frame the MHD models are similar.  In these two cases it seems that the dissipation structures are similar for incompressible and compressible MHD turbulence.

HD and MHD turbulence are similar over high order longitudinal scaling exponents measured in the global reference frame as well as for motions perpendicular to the local mean field.  In the local magnetic system the motions parallel to the local mean field for both MHD models are different from hydrodynamic motions.

\section{Fast, Slow, and Alfvenic Modes}

\begin{figure*}
\includegraphics{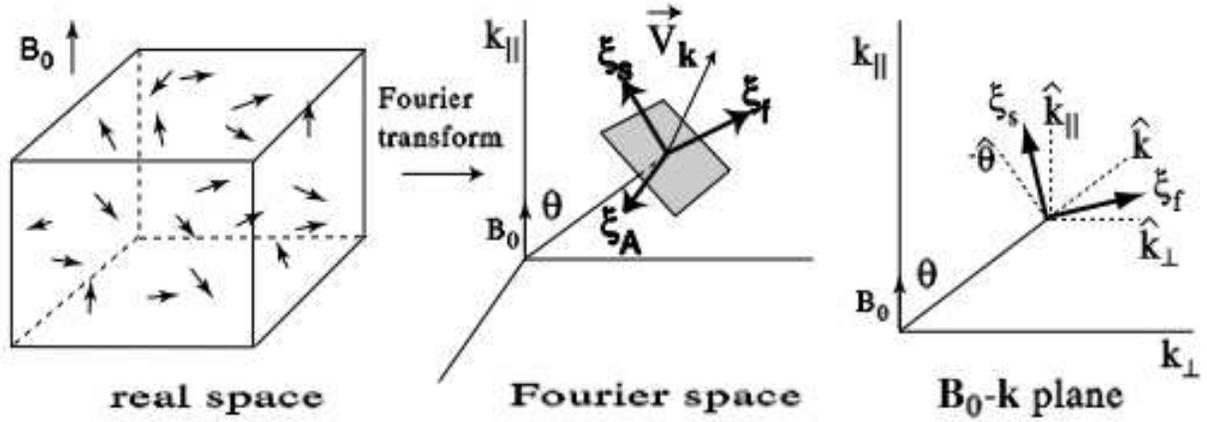}
\caption{Separation method. We separate Alfven, slow, and fast modes in Fourier space by projecting the velocity Fourier component ${\bf v_k}$ onto bases ${\bf \xi}_A$, ${\bf \xi}_s$, and ${\bf \xi}_f$, respectively.  Note that ${\bf \xi}_A = -\hat{\bf \varphi}$.  Slow basis ${\bf \xi}_s$ and fast basis ${\bf \xi}_f$ lie in the plane defined by ${\bf B}_0$ and ${\bf k}$.  Slow basis ${\bf \xi}_s$ lies between $-\hat{\bf \theta}$ and $\hat{\bf k}_{\|}$.  Fast basis ${\bf \xi}_f$ lies between $\hat{\bf k}$ and $\hat{\bf k}_{\perp}$ (from Cho \& Lazarian 2003).}
\label{decomp_explan}
\end{figure*}

\begin{figure*}
\includegraphics{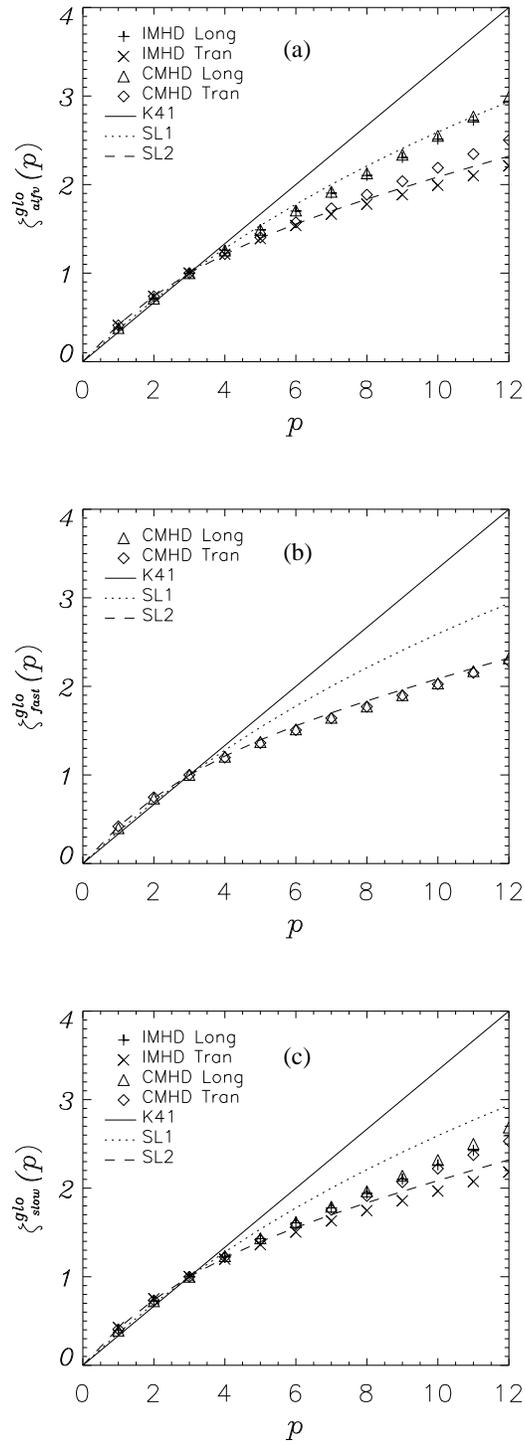}
\caption{Decomposed scaling exponents, normalized to the third order, as a function of the order of the structure functions of the velocity field in the global reference frame for the IMHD and CMHD models as well as K41, SL1, and SL2 (top - Alfvenic, middle - fast, and bottom - slow).
\label{decompses_glo}}
\end{figure*}

\begin{figure*}
\includegraphics{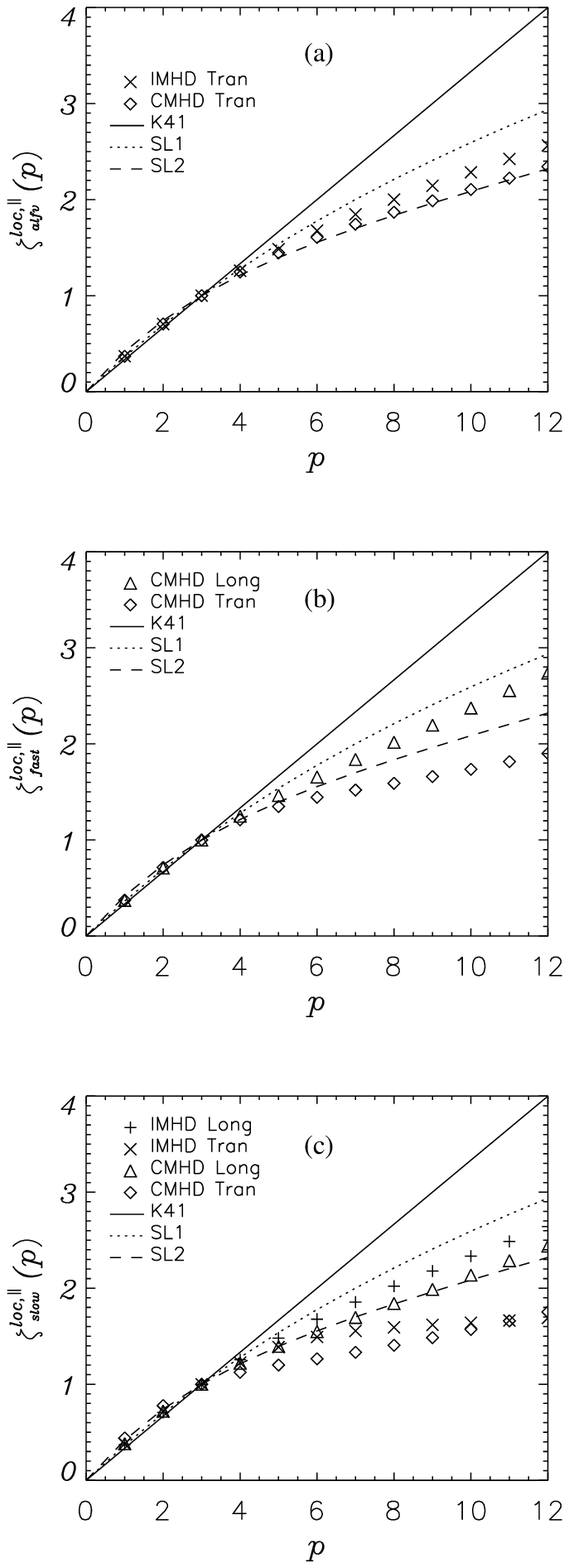}
\caption{Decomposed scaling exponents, normalized to the third order, as a function of the order of the structure functions in the local parallel reference frame for the IMHD and CMHD models as well as K41, SL1, and SL2 (top - Alfvenic, middle - fast, and bottom - slow).  
\label{decompses_loc_para}}
\end{figure*}

\begin{figure*}
\includegraphics{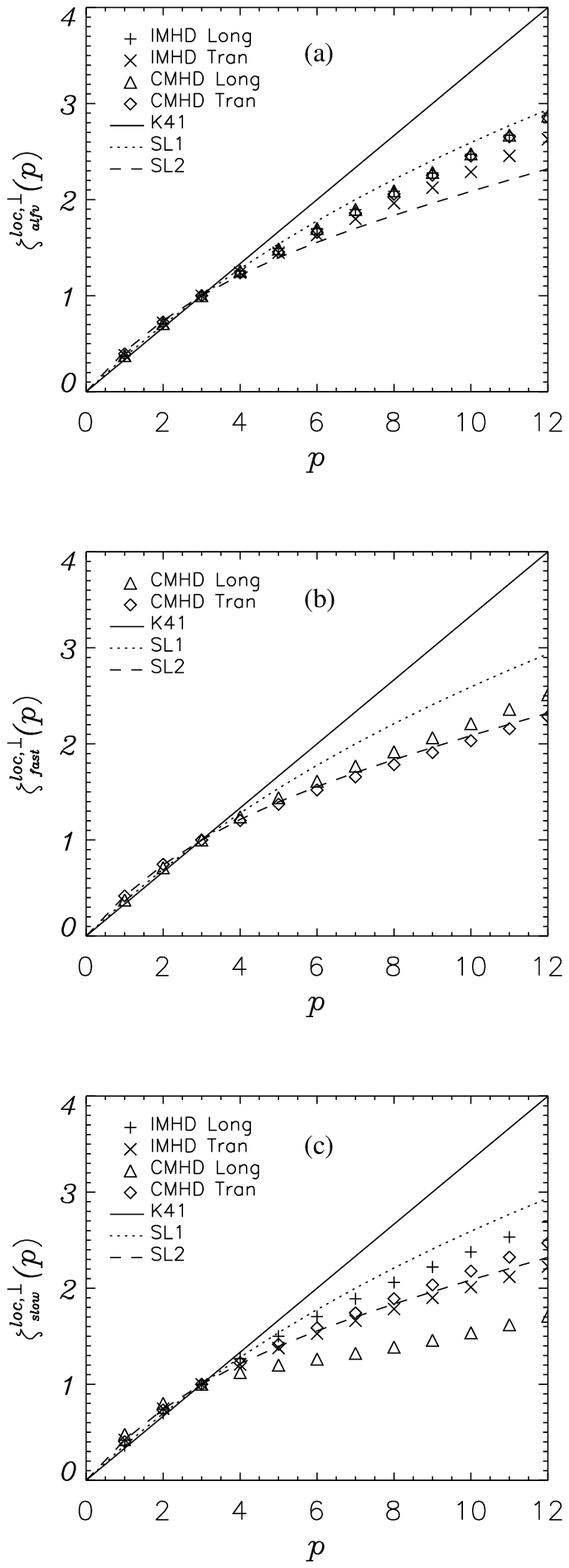}
\caption{Decomposed scaling exponents, normalized to the third order, as a function of the order of the structure functions in the local perpendicular reference frame for the IMHD and CMHD models as well as K41, SL1, and SL2 (top - Alfvenic, middle - fast, and bottom - slow).
\label{decompses_loc_perp}}
\end{figure*}

The compressible MHD equations support three types of linear waves or modes known as Alfvenic, fast, and slow.  In the incompressible limit the fast mode does not exist.  The velocity fields can be decomposed into these three components and scaling exponents for each component can be obtained, as explained in \cite{cho03b}\footnote{
   Note however that this decomposition method is an approximate
   one. It is well known that MHD allows for discontinuities and
   shocks. In the vicinity of these structures the wave decomposition
   method fails. We expect that these structures are more pronounced
   for higher sonic Mach number fluids. Therefore, a proper care
   should be taken for high Mach number fluids (see \cite{cl03}
   for an evaluation of the decomposition method for a high   
   Mach number case). In our current case, the Mach number is
   of order unity and, therefore, the discontinuous structures
   should not be very important.
}. Figure \ref{decomp_explan} gives a pictorial representation of the decomposition method \cite{cl03}.  Figures \ref{decompses_glo} through \ref{decompses_loc_perp} contain decomposed scaling exponents and are organized as follows (similar to Figures \ref{ses_global} through \ref{ses_loc_perp}).  In Figure \ref{decompses_glo} the global reference frame is used, in Figure \ref{decompses_loc_para} the local parallel reference frame is used, and in Figure \ref{decompses_loc_perp} the local perpendicular reference frame is used.  In each Figure the (a) is the Alfvenic mode, the (b) is the fast mode (containing only the CMHD model), and (c) is the slow mode.  For the IMHD model the field of the slow mode was calculated by subtracting the Alfvenic field from the total velocity field.  Each plot in Figures \ref{decompses_glo} through \ref{decompses_loc_perp} contains the scaling exponents derived from the longitudinal and transverse structure functions in order to expose the differences.  For the Alfv\'en mode in Figure \ref{decompses_loc_para}, given that in the direction parallel to $B_{ext}$ there is no Alfv\'en mode, the longitudinal components have no physical meaning and are not included.  

Beginning with Figure \ref{decompses_glo} in the global reference frame, for the Alfv\'en mode the longitudinal components of both models follow SL1 while the transverse follow closer to SL2.  For the the fast mode the CMHD longitudinal and transverse components are almost indistinguishable and follow SL2.  For the slow mode the transverse components of both models are more SL2-like.

In Figure \ref{decompses_loc_para} in the local parallel reference frame, for the Alfv\'en mode the CMHD transverse component follows SL2 quite well while the longitudinal component falls between SL1 and SL2.  For the fast mode the CMHD transverse component falls well below SL2 while the longitudinal component falls between SL1 and SL2.  For the slow mode the IMHD longitudinal component falls between SL1 and SL2 while the CMHD longitudinal component follows SL2 quite well.  The transverse components of both models fall well below SL2 and for the IMHD model the curve actually becomes flat.

In Figure \ref{decompses_loc_perp} in the local perpendicular reference frame, for the Alfv\'en mode all cases follow SL1 except the IMHD transverse component which falls between SL1 and SL2.  For the CMHD model the longitudinal and transverse components are indistinguishable.  For the fast mode both components of the CMHD model follow closer to SL2, the transverse component more so.  For the slow mode the IMHD longitudinal component follows closer to SL1 while the transverse component follows SL2 fairly well.  For the CMHD model this is the only case where the transverse component is significantly greater than the longitudinal component.  The transverse component follows SL2 while the longitudinal component falls well below SL2.

There are three cases in which scaling exponents fall well below SL2, namely the CMHD longitudinal component of $\zeta^{loc, \|}_{fast}(p)$, the transverse component for both models of $\zeta^{loc, \|}_{slow}(p)$, and the CMHD longitudinal component of $\zeta^{loc, \bot}_{slow}(p)$, suggesting that intermittency is greater in these cases.  

Also, $\zeta_L(p)\geq\zeta_T(p)$ for all cases except for the CMHD model of $\zeta^{loc, \bot}_{slow}(p)$.  This is not reflected in Figure \ref{ses_loc_perp} where for the CMHD model $\zeta_L(p)>\zeta_T(p)$.  This suggests that the amplitude of the slow mode for this model is small compared to the amplitude of the Alfvenic and fast modes.  The two cases where $\zeta_L(p)=\zeta_T(p)$ are for the CMHD models of $\zeta^{glo}_{fast}(p)$ and $\zeta^{loc, \bot}_{alfv}(p)$.  

In agreement with the above-mentioned property, $\zeta_L(p)\geq\zeta_T(p)$, it seems that the longitudinal components are more likely to be SL1-like while the transverse components are more likely to be SL2-like.  This is the case for $\zeta^{glo}_{alfv}(p)$ and $\zeta^{loc, \bot}_{slow}(p)$, which is in direct agreement with the conspicuous cases mentioned in the discussion of Figures \ref{ses_global} through \ref{ses_loc_perp}.  Combining the results of Figures \ref{ses_global} and \ref{decompses_glo} it seems that in the global reference frame the Alfvenic mode is mostly responsible for the fact that the longitudinal components of both MHD models follow SL1.  For the IMHD model both the Alfvenic and slow modes seem to contribute significantly to the fact that the transverse components appear to follow SL2.  Combining the results of Figures \ref{ses_loc_perp} and \ref{decompses_loc_perp} it seems that in the local perpendicular reference frame once again the Alfvenic mode is mostly responsible for the fact that the longitudinal components of both MHD models follow SL1.  For the IMHD model in this case the slow mode seems to contribute most to the fact that the transverse components appear to follow SL2.  

It is also true that for the decomposed scaling exponents of the two MHD models the longitudinal components are more alike than the transverse components, in every case except for $\zeta^{loc, \bot}_{slow}(p)$.  This makes sense in the local parallel reference frame where there is a bias in that the longitudinal components are chosen in the direction of the local mean field while the transverse components are chosen randomly to be perpendicular to the longitudinal components, but this observation also holds in the global reference frame where there is no bias because the longitudinal components are chosen randomly.  If the dissipation structures are similar for compressible and incompressible MHD then it is possible that the longitudinal scaling exponents would do a better job of probing this.

\section{Discussion and Summary}

In this paper turbulence in an incompressible fluid with and without a magnetic field as well as moderately compressible MHD turbulence are considered.  We summarize our findings as follows:  
\begin{enumerate}
\item Substantial similarity was shown between the spectra for all three models.
\item The PDF's of the velocity components perpendicular to the external magnetic field are IHD-like while those parallel to the field have a significantly smaller range of velocities.  The PDF's of the transverse velocity increments for the MHD models decay slower than the IHD model. 
\item The structure functions of the MHD models in the global frame and in the plane perpendicular to the local mean field are IHD-like in that the values of the transverse structure functions are greater than those of the longitudinal.  In these two frames the shapes of the structure functions of the IMHD model are more IHD-like than the CMHD model. 
\item The similarity of HD, IMHD, and CMHD turbulence persists over high order longitudinal structure function scaling exponents measured in the global reference frame as well as for motions perpendicular to the local mean field.  In these two frames the longitudinal scaling exponents of both MHD models seem to follow the predictions for one dimensional IHD-like dissipations structures while the transverse scaling exponents of the IMHD model seems to follow the predictions for two dimensional IMHD-like dissipation structures.  In the local magnetic system the motions parallel to the local mean field for both MHD models are different from hydrodynamic motions.
\item In the global reference frame and for motions perpendicular to the local mean field the Alfvenic mode is mostly responsible for the fact that the longitudinal components of both MHD models follow the IHD model and the IHD theoretical dissipation structure predictions.  In the global reference frame for the IMHD model both the Alfvenic and slow modes seem to contribute significantly to the fact that the transverse components follow the IMHD theoretical predictions.  For motions perpendicular to the magnetic field for the IMHD model the slow mode seems to contribute most to the fact that the transverse component seems to follow the IMHD theoretical predictions.  It is also true that for the decomposed scaling exponents of the two MHD models the longitudinal components are more alike than the transverse components, in every case except for the slow mode in the plane perpendicular to the local mean field.  
\end{enumerate}

\begin{acknowledgements}
The research of Nicholas Hall, Grzegorz Kowal, and Alex Lazarian was supported by the NSF grant AST0307869, Vilas Professorship Award and the Center for Magnetic Self-Organization in Astrophysical and Laboratory Plasmas.  Jungyeon Cho was supported by the Korea Research Foundation grant  KRF-2006-331-C00136.
\end{acknowledgements}

\clearpage

\end{document}